\documentclass[12pt,preprint]{aastex}
\usepackage{natbib}

\bibpunct{(}{)}{;}{a}{}{,}
\def\Ha{\textsc{H$\alpha$}}
\def\NII{\textsc{[N\,II]}$\lambda$6584}
\def\SII{\textsc{[S\,II]}$\lambda$$\lambda$6717, 6731}
\def\HII{\textsc{H\,II}}

\slugcomment{ }

\shorttitle{Wavelength     calibration      for     OSIRIS/GTC TFs}
\shortauthors{M\'endez-Abreu et al.}

\begin{document}


\title{Wavelength     calibration      for
     OSIRIS/GTC\altaffilmark{*} tunable      filters}


\author{J.~M\'endez-Abreu\altaffilmark{1,2}, 
           J. S\'anchez Almeida\altaffilmark{1,2},
           C. Mu\~noz-Tu\~n\'on\altaffilmark{1,2},
           J. M. Rodr\'iguez-Espinosa\altaffilmark{1,2},
           J. A. L. Aguerri\altaffilmark{1,2}, 
           D. Rosa Gonz\'alez \altaffilmark{3},
           Y. D. Mayya \altaffilmark{3},
           O. Vega \altaffilmark{3},
           R. Terlevich \altaffilmark{3,4},
           E. Terlevich\altaffilmark{3,4},
           E. Bertone \altaffilmark{3},
           L. H. Rodr\'iguez-Merino \altaffilmark{3}}

\altaffiltext{*}{This  work is based  on observations  made with
       the GTC operated on the island  of La Palma by Grantecan in the
       Spanish Observatorio del Roque de los Muchachos.}
\altaffiltext{1}{Instituto de Astrof\'\i sica de Canarias, E-38205 La Laguna, Tenerife, Spain.}
\altaffiltext{2}{Departamento de Astrof\'\i sica, Universidad de La Laguna, Tenerife, Spain.}
\altaffiltext{3}{Instituto Nacional de Astrof\'isica, \'Optica y Electr\'onica, Tonantzintla, Puebla, C.P. 72840, Mexico.}
\altaffiltext{4}{Visiting Fellow, IoA, Cambridge, UK}


\begin{abstract}
OSIRIS  (Optical  System for  Imaging  and  low Resolution  Integrated
Spectroscopy)  is the first  light instrument  of the  Gran Telescopio
Canarias (GTC).  It provides a flexible and competitive tunable filter
(TF).  Since  it is based  on a Fabry-Perot interferometer  working in
collimated  beam,  the  TF  transmission  wavelength  depends  on  the
position of the target with  respect to the optical axis.  This effect
is non-negligible and must be accounted for in the data reduction. Our
paper  establishes a  wavelength calibration  for OSIRIS  TF  with the
accuracy required  for spectrophotometric measurements  using the full
field  of  view  (FOV)  of  the  instrument.   The  variation  of  the
transmission wavelength $\lambda(R)$ across  the FOV is well described
by     $\lambda(R)=\lambda(0)/\sqrt{1+\left(R/f_2\right)^2}$,    where
$\lambda(0)$ is  the central  wavelength, $R$ represents  the physical
distance  from the  optical axis,  and $f_2=185.70\pm0.17\,$mm  is the
effective  focal  length  of  the  camera lens.   This  new  empirical
calibration yields an accuracy better than $1$\,\AA\ across the entire
OSIRIS   FOV  ($\sim$8\arcmin$\times$8\arcmin),   provided   that  the
position of the  optical axis is known within  45~$\mu$m ($\equiv$ 1.5
binned  pixels).  We suggest  a  calibration  protocol  to grant  such
precision over  long periods, upon re-alignment of  OSIRIS optics, and
in  different wavelength  ranges.  This  calibration differs  from the
calibration in OSIRIS manual  which, nonetheless, provides an accuracy
$\lesssim1$\AA\, for $R\lesssim 2\arcmin$.
\end{abstract}



\section{Introduction}
\label{sec:introduction}

Tunable  filter  (TF)  instruments  are becoming  standard  tools  for
mapping  physical properties  in extended  astronomical  sources.  Most  modern   large  telescopes  have  TFs   based  on  Fabry-Perot
  interferometers (FPI; or etalons) such as the 11\,m Southern African
  Large Telescope (SALT; Rangwala  et al.  2008) or the Magellan-Baade
  6.5 m  telescope \citep{veilleux10}.  The tunable  imaging era began
  with the  instrument of \citet{athertonreay81} and  many others have
  further developed  the technique \citep{blandtully89,brown94}.  The
  current instrumental  design is mainly  based on the  Taurus Tunable
  Filter  (Bland-Hawthorn  \&  Jones  1998) formerly  mounted  at  the
  Anglo-Australian  Telescope  and  afterward  moved  to  the  William
  Herschel Telescope at La Palma.   The power of these instruments to
perform   spectrophotometric  studies  is   clear.   They   allow  for
spectrophotometry with an unlimited  number of wavelengths that can be
scanned within a  wavelength range.  These etalon based  TFs provide a
central monochromatic field (MF).  Outside  the MF, and due to the use
of  etalons  in collimated  beam,  there  is  a significant  shift  of
wavelength  that   depends  on  the  distance  to   the  center.   The
instruments are designed to provide the largest possible MF, the wider
possible  wavelength  range, and  the  narrower possible  transmission
band-pass.   Besides,  a proper  calibration  corrects the  wavelength
displacement with  radius, thus recovering  much of the  field outside
the MF to be used as an extended MF.

Among the large  facilities hosting such instruments is  the 10~m Gran
Telescopio  Canarias (GTC),  located  at the  Roque  de los  Muchachos
Observatory, that started its operations in the first semester of 2009
with OSIRIS \citep{cepa05,cepa10} as the first light instrument. OSIRIS
stands for  "Optical System for Imaging and  low Resolution Integrated
Spectroscopy".   Besides  conventional  imaging and  spectroscopy,  it
provides the additional  capability of a TF observing  mode working in
the visible.  It  covers the wavelength range 0.365-1.0  $\mu$m with a
nominal  FOV of  7.8\arcmin$\times8.5$\arcmin.   Further commissioning
tests     establish      the     real     un-vignetted      FOV     as
$7.8\arcmin\times7.8\arcmin$       (see      OSIRIS      commissioning
webpage\footnote{http://www.gtc.iac.es/en/pa\-ges\-/\-ins\-tru\-men\-ta\-tion\-/\-o\-si\-ris\-/\-da\-ta\--\-co\-mmi\-ssio\-ning.php}).
OSIRIS specifications are extensively described in the manual provided
by      the       instrument      team      (OSIRIS       TF      user
manual\footnote{http://www.gtc.iac.es/en/pages/\-ins\-tru\-men\-ta\-tion\-/\-o\-si\-ris\-/\-o\-si\-ris2\-.php\-$\#$U\-se\-ful\-$\_$Do\-cuments;
  current version dated June 28, 2009} -- this document is referred to
along the paper  as OSIRIS {\em manual}).  In  particular, it provides
the wavelength calibration procedure for radii outside the MF.

The OSIRIS FPI is an ideal  instrument to map emission line regions in
galaxies.   It provides  a  high throughput  with narrow  transmission
bandpass.   The central wavelength  is adjustable  to account  for the
galaxy recession velocities, and the field of view (FOV) is reasonably
large  with  a  small  pixel  scale  (0\farcs125).   These  properties
together with  the collecting power of  a 10 m  telescope, designed to
get the best possible natural image quality, prompted us to design the
Local                          Universe                         Survey
(LUS\footnote{http://www.inaoep.mx/$~$gtc$-$lus/}), a research program
optimized for OSIRIS.  LUS aims at studying the star-formation history
of a  complete sample of nearby  galaxies.  We planned  to obtain high
signal-to-noise,  high angular  resolution,  narrow band  maps of  all
galaxies  inside a  volume  of  3.5 Mpc  radius.   All irregulars  and
spirals inside a volume of 11  Mpc were also included, together with a
sample  of  Virgo cluster  galaxies.   These  observations suffice  to
produce  a detailed  unique  description of  the  gaseous and  stellar
content  of  nearby  galaxies.    LUS  has  been  defined  within  the
ESTALLIDOS\footnote{http://estallidos.iac.es/estallidos/Estallidos.jsp}
project  in  close  collaboration   with  the  Instituto  Nacional  de
Astrof\'isica,  \'Optica  y   Electr\'onica  in  Mexico.   Significant
preparatory work was carried out  in order to prove the feasibility of
LUS;  using the  OSIRIS  features  described in  the  manual, we  made
simulations to  estimate the exposure  times, to choose  precisely the
filters (broad band and TF),  and to optimize the observing procedure.
Before submitting LUS as an  ESO-GTC large program, we decided to test
our simulations and procedures with real data.

The object selected  to calibrate  the photometric  accuracy was
  M101, an  extended and well-known nearby galaxy  (7.38~Mpc, Rizzi et
  al. 2007), that  exceeds the size of OSIRIS  FOV.  Observations were
  carried out to cover a large  region of the southwestern part of the
  galaxy,  containing also the  well-known giant  extragalactic \HII\,
  region NGC 5447.   The central wavelengths of the  TF were chosen to
  map  emission lines  such as  \Ha\, or  the sulphur  doublet (\SII),
  which  allow  to derive  star-formation  rates  (SFRs) and  electron
  densities  of  the  star-forming  regions.   Even  if  not  optimal,
  observations were made using the narrowest bandpass of FWHM 18 \AA\,
  available at the time.  The precise central wavelength of the filter
  needed   to  separate   \NII\  from   \Ha\  was   also   tested,  as
  distinguishing  the  two  lines  is  critical  to  be  able  to  use
  diagnostic  diagrams  like BPT  \citep{baldwin81}.   The basic  data
  reduction  was  carried  out  using  our  own  pipeline.   Then,  we
  reconstructed monochromatic emission line images taking into account
  the  wavelength calibration provided  in OSIRIS  documentation. From
  this analysis,  we found  that \Ha\, fluxes  and SFRs  were globally
  consistent with  values from the literature.   However, the emission
  line ratios for both \NII\  over \Ha\ and the sulphur doublet (\SII)
  were far from  any observed value and whatever  \HII\, region model.
  In an  attempt to identify the  problem, we started  from scratch by
  using spectral  calibration lamps  to test wavelength  shifts beyond
  the MF.  The result of our calibration is the content of this paper.
  We  have figured  out that  the  prescriptions given  in the  manual
  suffice for the  MF, while the new calibration  in this paper allows
  for   using   OSIRIS   in   its   full   capability.    It   permits
  spectrophotometry in the full FOV with a wavelength accuracy as good
  as that  in the  MF.  In  our view OSIRIS  is an  extremely powerful
  instrument which  could be seriously limited  if only the  MF can be
  used  for  spectrophotometry.  Although  our  work  was designed  to
  recalibrate our  fields and being able to  recover reliable emission
  line fluxes over the entire OSIRIS FOV, it is of interest beyond the
  original scope, and may broaden the community of OSIRIS users.

The  paper  is  organized  as follows:  in  \S~\ref{sec:optics}  we
introduce the  basic optical concepts required to  understand both the
nature    of   the    problem   and    the    calibration   procedure.
\S~\ref{sec:urge}  shows how  the  original wavelength  calibration
needs refinement to provide an accuracy better than 1\AA\, in the full
FOV.  Based  on the data  introduced in \S~\ref{sec:data},  the new
calibration    is    worked    out   in    \S~\ref{sec:cal}.     In
\S~\ref{sec:protocol}   we  recommend   a   wavelength  calibration
protocol that goes  all the way from a detailed  test of the long-term
stability of the system, to a black-box recipe for urgent calibration.
The     conclusions    of     the    work     are     summarized    in
\S~\ref{sec:conclusions}.

\section{Basics of OSIRIS/GTC tunable filter optics}
\label{sec:optics}

\begin{figure*}[!ht]
\centering
\includegraphics[angle=-90,scale=0.5]{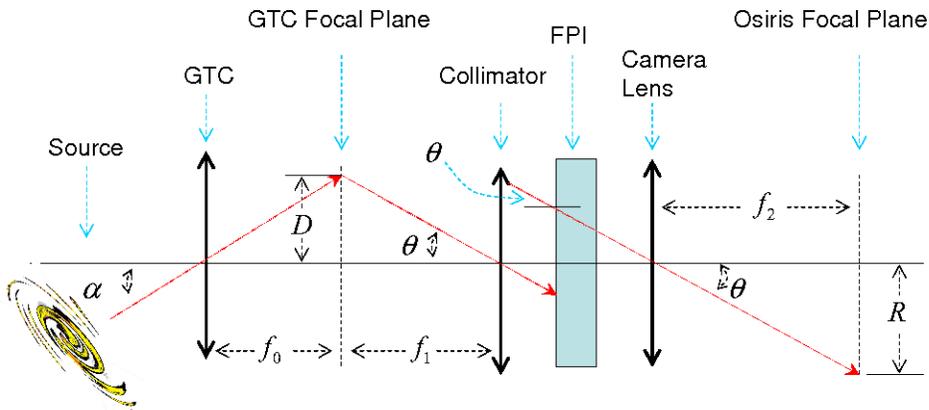}
\caption{Schematic of the OSIRIS/GTC optical layout.  A light beam, in
  red,  crosses  the  system  from  left to  right,  finding  the  GTC
  telescope,  the  collimator,  the  Fabry-Perot  interferometer,  the
  camera, and  OSIRIS focal plane.  The different  symbols are defined
  in the main text.}
\label{fig:layout}
\end{figure*}
%

In  essence, the optical  layout of  OSIRIS/GTC TFs  is made  of three
lenses  and a FPI,  as sketched  in Fig.~\ref{fig:layout}.   The first
element,  of effective  focal length  $f_0$, represents  the telescope
(GTC).  Then a  lens, of effective focal length  $f_1$, collimates the
light coming from the GTC focal  plane, so that the FPI is illuminated
in collimated beam.  As  represented in Fig.~\ref{fig:layout}, a point
in the  focal plane of GTC, at  a distance $D$ from  the optical axis,
enters the FPI interferometer with an angle $\theta$.  Then
%
\begin{equation}
D=f_1\tan\theta=f_0\tan\alpha\simeq f_0\alpha,
\label{1steq}
\end{equation}
%
where $\alpha$ represents the angular  distance to the optical axis of
the source on the sky.  The wavelength transmitted by a FPI depends on
the incidence angle as
\begin{equation}
\lambda(\theta)=\lambda(0)\cos\theta\,,
\label{fpieq}
\end{equation}
%
where $\lambda(0)$ is the central wavelength (e.g., Born \& Wolf 1980;
Beckers 1998; Veilleux  et al. 2010).  The final  optical element, the
camera, has an effective focal length $f_2$, so that,
\begin{equation}
R=f_2\tan\theta,
\label{cameraeq}
\end{equation}
%
where $R$  is the position  of the source  on  OSIRIS  focal plane.
Combining  Eqs.~(\ref{fpieq}) and  (\ref{cameraeq}), one  gets  the TF
wavelength calibration  equation in terms of the  physical distance to
the optical axis $R$. It is given by
%
\begin{equation}
\lambda(R)=\frac{\lambda(0)}{\sqrt{1+\left(\frac{R}{f_2}\right)^2}},
\label{eq:caleq1}
\end{equation}
%
or equivalently,  employing Eq.~(\ref{1steq}), it can  be expressed in
terms of the angular distance projected on the sky
%
\begin{equation}
\lambda(\alpha)=\frac{\lambda(0)}{\sqrt{1+\left(\alpha\, \frac{f_0}{f_1}\right)^2}}.
\label{eq:caleq2}
\end{equation}
%
If  $\alpha$  is  small  enough,  then  Eq.~(\ref{eq:caleq2})  can  be
expanded to first order giving rise to
%
\begin{eqnarray}
\lambda(\alpha) &\simeq& \lambda(0)\,\big(1-C\,\alpha^2\big),\label{eq:approx1}\\
C&=&\frac{f_0^2}{2f_1^2},\nonumber
\end{eqnarray}
%
and equivalently in terms of the physical distance
%
\begin{eqnarray}
\lambda(R)&\simeq& \lambda(0)\,\big(1-C^*\,R^2\big),\label{eq:approx2}\\
C^*&=&\frac{1}{2f_2^2}\nonumber.
\end{eqnarray}

The  plate scale  on OSIRIS  focal plane  relates the  effective focal
lengths of  the three optical  elements.  Combining Eqs.~(\ref{1steq})
and (\ref{cameraeq}), the plate scale $S$ turns out to be
%
\begin{equation}
S={d\alpha\over{dR}}={{f_1}\over{f_0f_2}},
\label{eq:scale}
\end{equation}
%
which we use to link the two calibration constants, 
\begin{equation}
C^*=C\,S^{2}.
\label{eq:linking}
\end{equation}

OSIRIS   manual    employs   Eq.~(\ref{eq:approx1})   for   wavelength
calibration, therefore, it assumes the  FOV to be small enough for the
first  order  expansion to  hold.   In  addition,  the calibration  is
performed in angular coordinates on  the sky, so that uncertainties in
the plate  scale affect the  wavelength calibration. We  propose using
Eq.~(\ref{eq:caleq1})  instead.  It  makes  the calibration  sensitive
only to the uncertainties of the focal length of the camera $f_2$, and
it considers all  the terms of the expansion.  As we  show in the next
section, high  order terms are  needed to calibrate the  field outside
the MF.

\begin{deluxetable}{lcll}
\tabletypesize{\scriptsize}
\rotate
\tablecaption{Summary  of  the  basic  optical  parameters  for  OSIRIS/GTC wavelength calibration.\label{tab:summary}}
\tablehead{
\colhead{Physical Parameter}           & \colhead{Symbol}                   & \colhead{Nominal Value}                     & \colhead{Best Estimate}                                  
}
\startdata
GTC eff. focal length        & $f_0$                       & 170\,m $[{\rm a}]$          &  165.3 $\pm$ 0.2\,m   $[{\rm h}]$  \\
Collimator eff. focal length & $f_1$                       & 1.240\,m $[{\rm b}]$             & 1.275 $\pm$ 0.001\,m      $[{\rm i}]$                          \\
Camera eff. focal length     & $f_2$                       & 181\,mm $[{\rm b,c}]$            & 185.70 $\pm$ 0.17\,mm                              \\
Calibration constant         & $C=f_0^2/(2\,f_1^2)$    & $7.9520\,\times\,10^{-4}$arcmin$^{-2}$ $[{\rm d}]$ & (7.517 $\pm$ 0.014)$\,\times\, 10^{-4} $arcmin$^{-2}$$[{\rm j}]$ \\
Calibration constant         & $C^*=1/(2\,f_2^2)$          &  15.262 m$^{-2}$              & 14.499 $\pm$ 0.027 m$^{-2}$                         \\
                             &                             &  \hspace{1cm}    ---                     & (3.2624 $\pm$ 0.0057) $\times\, 10^{-9}$ pixel$^{-2}$\\
OSIRIS pixel size    (1$\times$1)        &        & 15\,$\mu$m $[{\rm a,b,c,d}]$  &                 \hspace{1cm}    ---            \\
OSIRIS pixel size    (2$\times$2)        &        & 30\,$\mu$m                    &                 \hspace{1cm}    ---            \\
OSIRIS plate scale   (1$\times$1)        & $S=f_1/(f_0f_2)$            &  0\farcs125\,pixel$^{-1}$ $[{\rm a,b,d,e}]$   &  0\farcs127\,pixel$^{-1}$ $[{\rm k}]$\\
OSIRIS plate scale   (2$\times$2)        &     &  0\farcs250\,pixel$^{-1}$    &       \hspace{1cm}    ---    \\
OSIRIS/GTC eff. focal length & $f_0f_2/f_1$                 & 24.752\,m  $[{\rm b,g}]$    &              \hspace{1cm}    ---                \\
OSIRIS center position  (2$\times$2)    & (x$_{\rm c1}$, y$_{\rm c1}$)   & (1059$\pm$ 2, 983$\pm$ 2) $[{\rm f}]$                & (1053.6 $\pm$ 2.2,979.5 $\pm$ 1.8)\\
                                        & (x$_{\rm c2}$, y$_{\rm c2}$)   & (-1$\pm$ 2, 981$\pm$ 2)  $[{\rm f}]$                 & (-9.8 $\pm$ 2.7,975.8 $\pm$ 1.1)\\
\enddata
\tablenotetext{a}{Cobos et al. (2002).}
\tablenotetext{b}{OSIRIS       project      webpage       {\tt http://www.iac.es/project/OSIRIS/}.}
\tablenotetext{c}{Cepa et al. (2005).}
\tablenotetext{d}{Value included in OSIRIS/GTC TF  user manual, and derived from the quoted nominal values for $f_0$ and $f_1$.}
\tablenotetext{e}{Cepa (2010).}
\tablenotetext{f}{OSIRIS/GTC TF webpage {\tt http://www.gtc.iac.es/en/pages/instrumentation/osiris.php\#Tunable\_Filters.}}
\tablenotetext{g}{It does not agree with the other nominal values -- it is obtained for $f_0=169.57$\,m.}
\tablenotetext{h}{Value calculated using Eq.~(\ref{eq:scale}) assuming the nominal values for $S$ and $f_1$, and the best estimate for $f_2$.}
\tablenotetext{i}{Value calculated using Eq.~(\ref{eq:scale}) assuming the nominal values for $S$ and $f_0$, and the best estimate for $f_2$.}
\tablenotetext{j}{Value calculated assuming the $S=$0\farcs125\,pixel$^{-1}$ nominal value. For other plate scales, see Eq.~(\ref{eq:calconst}).}
\tablenotetext{k}{Obtained from astrometry and in agreement with laboratory tests.} 
\end{deluxetable}


\section{Motivation for an updated calibration}
\label{sec:urge}

Inconsistencies  in emission  line  ratios prompted  us to  reconsider
OSIRIS  wavelength  calibration  (\S~\ref{sec:introduction}).   The
importance  of  this effect  can  be  disclosed  and quantified  using
monochromatic images of known wavelength, as those provided by the GTC
wavelength calibration unit.  When  the instrument is illuminated with
a  monochromatic beam  of known  wavelength $\lambda_{\rm  line}$, the
ring  appearing in the  focal plane  traces where  OSIRIS transmission
matches $\lambda_{\rm line}$ (see Fig.~\ref{fig:lamp}).  In principle,
the  position of  the optical  axis and  the tuned  central wavelength
$\lambda(0)$  are  given,  therefore,   one  can  easily  measure  the
difference between the true  wavelength $\lambda_{\rm line}$, and that
provided    by   the    nominal   calibration    in    OSIRIS   manual
(Eq.~(\ref{eq:approx1})).  If  the ring is  observed to have  a radius
$R$ (Fig.~\ref{fig:lamp}), then the error in wavelength is given by

\begin{equation}
\Delta\lambda=\lambda_{\rm line}-\lambda(R)=
\lambda_{\rm line}-\lambda(0)\,\big(1-C\,S^2\,R^2\big),
\label{eq:needfor}
\end{equation}
where  $C=7.9520\times10^{-4} {\rm  arcmin}^{-2}$  is the  calibration
constant appearing in  the manual, and $S$ is  the nominal plate scale
($0\farcs125\,$ pixel$^{-1}$;  see Table~\ref{tab:summary} with OSIRIS
properties).

The  black dots  in Fig.~\ref{fig:badcal}  show $\Delta\lambda$  for a
suite  of 17 monochromatic  images used  for calibration  (details are
given  in  \S~\ref{sec:data}). Note  that  the  error  in the  true
wavelength can be as large  as 6\,\AA\, at the FOV outskirts ($R\simeq
4$\arcmin;  see  Fig.~\ref{fig:badcal}).   If  rather than  using  the
approximate   Eq.~(\ref{eq:approx1}),   theoretical  wavelengths   are
computed using  the full  Eq.~(\ref{eq:caleq2}), then one  obtains the
triangles  in Fig.~\ref{fig:badcal}.   The errors  have  been somewhat
reduced, but they still remain  as large as 4\,\AA\, at 4\arcmin.  Two
main conclusions can be  drawn from Fig.~\ref{fig:badcal}.  First, the
original  calibration procedure does  not provide  correct wavelengths
within 1\,\AA\, when  $R> 2$\arcmin. Second, in order  to achieve such
an accuracy,  one has to  use the full  Eq.~(\ref{eq:caleq1}) combined
with a new  calibration constant more precise than  that inferred from
the nominal values of the focals.

\section{Data}
\label{sec:data}

\begin{figure}[!ht]
\centering
\includegraphics[width=0.7\textwidth]{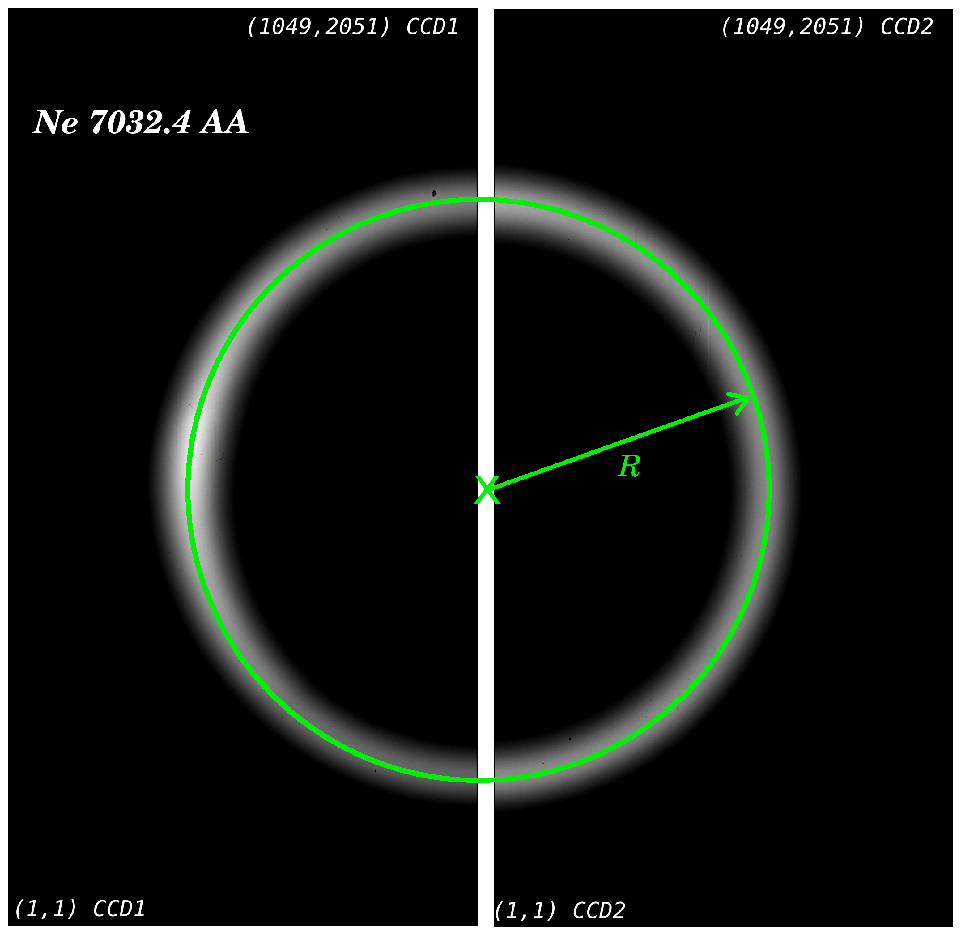}
\caption{Image  of a  NeI spectral  calibration lamp  illuminating the
  OSIRIS  detectors. It forms  a ring  which we  have outlined  with a
  green circle of radius $R$ centered at $X$. OSIRIS has been tuned to
  a central wavelength of 7070~\AA, whereas the NeI emission occurs at
  7032.4 \AA.  Binned coordinates (1,1) and (1049,2051) are also drawn
  on the image.}
\label{fig:lamp}
\end{figure}

\begin{figure}[!ht]
\centering
\includegraphics[width=0.7\textwidth]{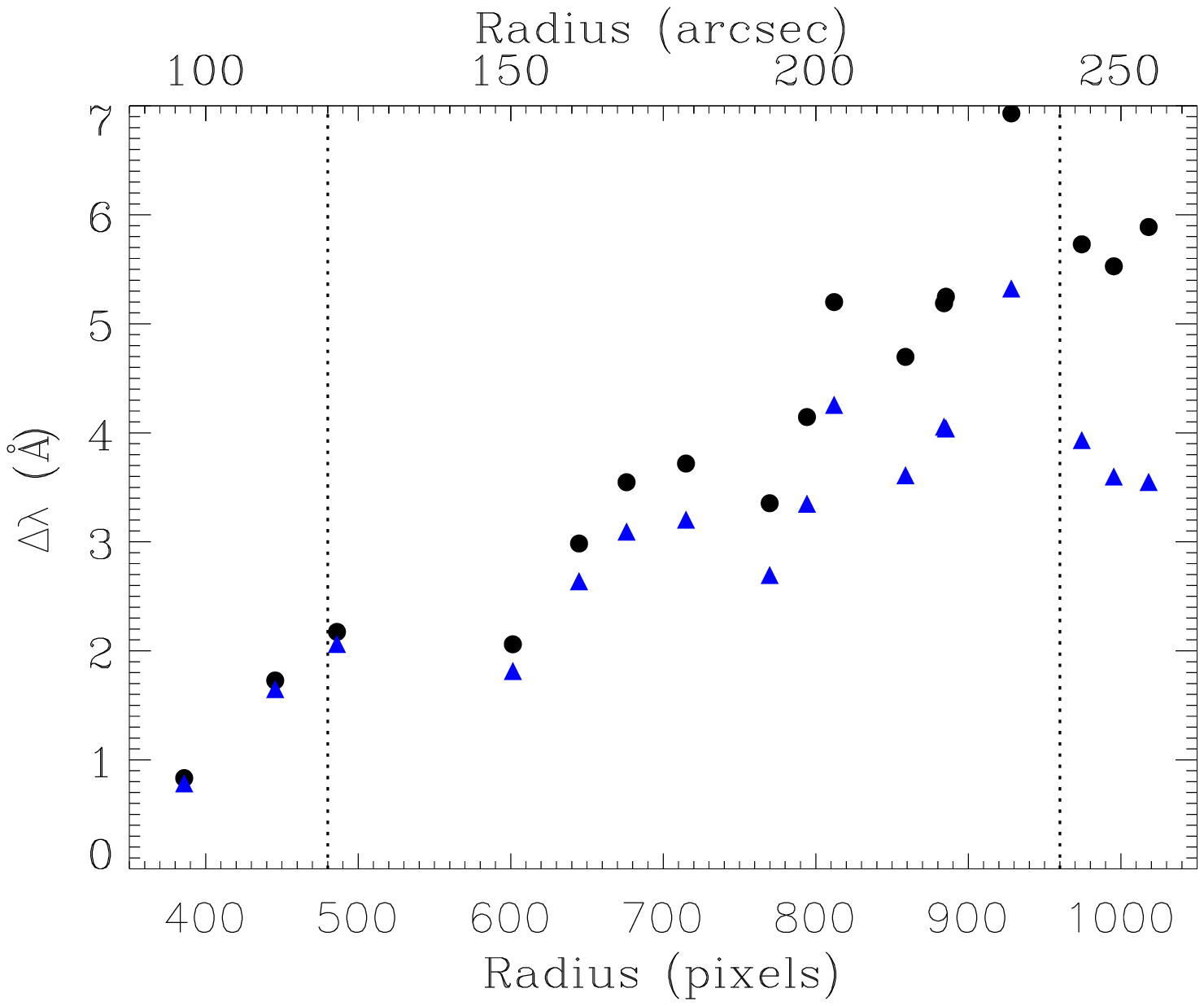}
\caption{Differences  between the  true wavelengths  and  that derived
  using the nominal OSIRIS calibration procedure.  Each point has been
  inferred from one image similar to that in Fig.~\ref{fig:lamp}.  The
  discrepancies  increase  with distance  from  the  optical axis  $R$
  ($\equiv  Radius$).   The  black  dots  correspond  to  the  nominal
  calibration based on the approximate Eq.~(\ref{eq:approx1}), whereas
  the blue  triangles used the  full Eq.~(\ref{eq:caleq2}).  Abscissas
  are given in pixels (bottom) and arcsec (top), the latter calculated
  using  the nominal  plate  scale.  Vertical  dotted  lines show  the
  radius  of  the  MF   ($\sim$2\arcmin)  and  the  non-vignetted  FOV
  ($\sim$4\arcmin).}
\label{fig:badcal}
\end{figure}

Figure~\ref{fig:badcal}   and   the    calibration   worked   out   in
\S~\ref{sec:cal}  are based  on a  sample of  17  monochromatic images
produced by  arc lamps from  the Instrument Calibration Module  of GTC
(ICM).  They were obtained in  two different runs and using the OSIRIS
red    arm.     Their    main    properties    are    summarized    in
Table~\ref{tab:data}.   The images  were taken  by the  GTC  staff who
kindly provided them upon request.   The OSIRIS FOV is imaged onto two
CCDs that  are separated by a  gap.  The distance between  the CCDs is
approximately 9\farcs2.  The  two OSIRIS science CCDs are  of type E2V
44-82, back  illuminated, with  2048 x 4096  pixels each when  using a
1$\times$1 binning.  OSIRIS was tuned to different central wavelengths
in order  to produce the  characteristic ring for  every monochromatic
line  (Fig.~\ref{fig:lamp}).  They  were obtained  using  the standard
mode  of operation  with a  binning 2$\times$2.   The separation
  between  any two consecutive  orders of  interference is  called the
  free spectral  range.  In order  to select the  desired transmission
  order of the FP a blocking  filter is required.  The filters used in
  this study are also shown in Table \ref{tab:data}.

In addition  to the above  monochromatic exposures, three  images with
strong sky emission rings were  also requested in order to compare the
wavelength calibration obtained  from the ICM and from  the sky.  They
are also listed in Table~\ref{tab:data}.

\begin{table}
\begin{center} 
\caption{Characteristics of the calibration data.}
\label{tab:data}
\begin{tabular}{c c c c c c}
\noalign{\smallskip}
\tableline\tableline
$\lambda(0)$ & $\lambda_{\rm line}$ & Species & TF FWHM & Blocking Filter & Observing\\
(\AA) & (\AA) & & (\AA) & (nm/nm) & Date \\
\tableline
6598.9 & 6532.9 &NeI& 18 & f657/35 &10-02-2011 \\
6630.0 & 6598.9 &NeI& 18 & f657/35 &10-02-2011 \\
6650.0 & 6598.9 &NeI& 18 & f657/35 &10-02-2011 \\
6950.0 & 6929.5 &NeI& 16 & f694/44 &03-03-2011 \\
6975.0 & 6929.5 &NeI& 16 & f694/44 &03-03-2011 \\
7000.0 & 6929.5 &NeI& 16 & f694/44 &03-03-2011 \\
7020.0 & 6929.5 &NeI& 16 & f694/44 &03-03-2011 \\
7050.0 & 7032.4 &NeI& 16 & f694/44 &03-03-2011 \\
7070.0 & 7032.4 &NeI& 16 & f709/45 &03-03-2011 \\
7090.0 & 7032.4 &NeI& 16 & f709/45 &03-03-2011 \\
7100.0 & 7032.4 &NeI& 16 & f709/45 &03-03-2011 \\
7120.0 & 7032.4 &NeI& 16 & f709/45 &03-03-2011 \\
7650.0 & 7635.1 &ArI& 18 & f770/50 &03-03-2011 \\
7680.0 & 7635.1 &ArI& 18 & f770/50 &03-03-2011 \\
7700.0 & 7635.1 &ArI& 18 & f770/50 &03-03-2011 \\
7720.0 & 7635.1 &ArI& 18 & f770/50 &03-03-2011 \\
7740.0 & 7635.1 &ArI& 18 & f770/50 &03-03-2011 \\
\tableline
\noalign{\smallskip}
6900.0 & 6834.4-6841.9 & telluric OH & 12 & f666/36 & 12-02-2011 \\
6925.0 & 6871.0-6880.9 & telluric OH & 12 & f666/36 & 12-02-2011 \\
6950.0 & 6864.0-6871.0 & telluric OH & 12 & f680/36 & 12-02-2011 \\
\tableline
\end{tabular}
\end{center}
\end{table}

\section{Wavelength calibration}
\label{sec:cal}

As   we  argue   in  \S~\ref{sec:urge},   improving   the  original
calibration  demands both  including all  the terms  neglected  in the
approximate  Eq.~(\ref{eq:approx1})  and  determining the  calibration
constants directly  from observations.   We measured $C^*$  because it
does not require  knowing the plate scale $S$,  therefore, we make the
calibration  insensitive  to  errors  in $S$.   Besides,  $C$  follows
directly from $C^*$ and $S$ via Eq.~(\ref{eq:linking}).

Calibrating an  image is assigning a  wavelength to each  point of the
FOV  using Eq.~(\ref{eq:caleq1}).   Three different  unknowns  must be
set, namely,  the {\em position  of the center}, origin  of distances,
the  {\em calibration  constant},  and the  {\em central  wavelength}.
This  section  describes a  precise  determination  of  the two  first
ingredients  using the  data described  in  \S~\ref{sec:data}.  The
central wavelength is determined within 1\,\AA\ as part of the regular
OSIRIS tuning protocol.  This claim is confirmed in \S~\ref{sec:cc}
since it also affects the empirical determination of $C^*$.  The three
unknowns have  uncertainties that determine  the calibration accuracy.
The propagation  of these uncertainties  into the final  wavelength is
studied in \S~\ref{sec:budget}.
%

\subsection{TF optical center}
\label{sec:tfcenter}

The nominal  TF optical  center is located  in the pixel  (2118, 1966)
from the  pixel (1,1)  of CCD1, including  the 50 pixels  of overscan.
Thus,   assuming   the   nominal   gap   between  the   CCDs   of   72
pixels\footnote{http://www.gtc.iac.es/en/pages/\-ins\-tru\-men\-ta\-tion\-/\-o\-si\-ris\-.php\-$\#$\-Tu\-na\-ble\-$\_$\-Fil\-ters.},
the  center   lies  within  the  gap   of  the  two   CCDs  (see  Fig.
\ref{fig:lamp}).  As  pointed out before,  our images have  been taken
with the  standard 2$\times$2 binning.   Therefore, we might  expect a
center     positioned    at     binned    pixels     (1059,983) --    see
Table~\ref{tab:summary}.   Taking  into  account  that  the  distance
between  the CCDs  is $\sim$36  binned pixels,  the TF  optical center
should be placed 1 pixel to the left of column 1 of CCD2.

We have  measured the optical  center for our calibration  lamps using
the  following procedure:  first we  generate  a grid  of possible  TF
optical pixel centers  ($x_{\rm trial}$,$y_{\rm trial}$) independently
for both CCDs.  Since the gap  between CCDs is not constant (it varies
from top  to bottom  due to both  a rotation  and a 2  pixels vertical
shift  of  CCD2  with  respect  to CCD1),  the  solution  of  assuming
independent centers  bypasses the  difficulty of knowing  the relative
position of the two CCDs.  Then, for every trial position on each CCD,
the calibration lamp  image was divided in 8  regions of 45$^\circ$ (4
in each CCD).  Each region is azimuthally averaged and the distance of
the  ring to the  assumed center  is measured.   Due to  both possible
irregularities in the  optical system and the offset  between the true
and trial centers, the measurements of the ring radii for each region
are not  the same.  We assign the  best center from our  grid of trial
values  looking for  the value  that, providing  the same  mean center
within the errors for both CCDs, minimizes the scatter among the radii
of  the eight  different sectors.   If the  monochromatic  images were
noiseless  and   the  ring  perfectly   circular,  our  center-finding
algorithm would provide the  correct value.  In practice, however, the
procedure  has  some errors  that  we  estimate  by comparing  centers
obtained from different rings.

\begin{figure}[!ht]
\centering
\includegraphics[width=0.7\textwidth]{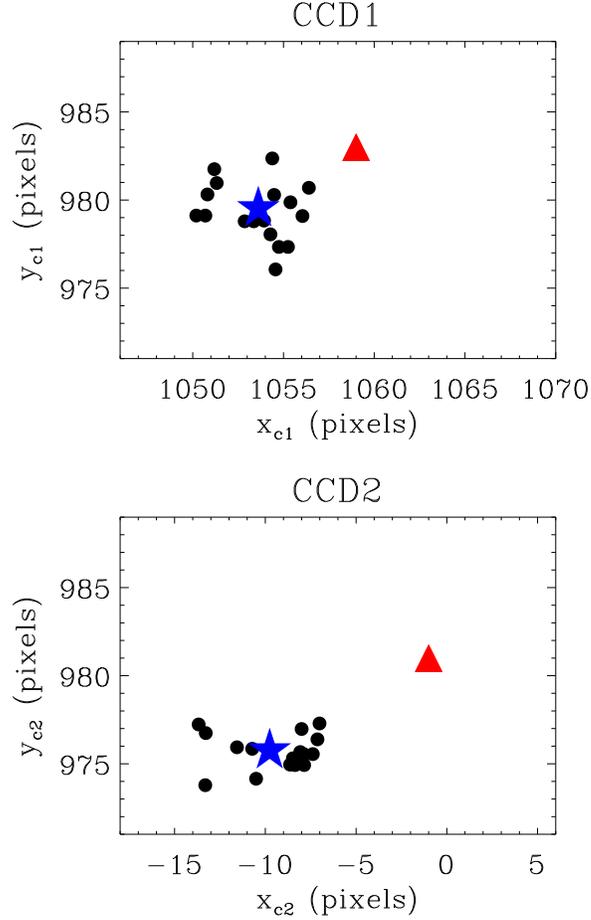}
\caption{  OSIRIS  optical  centers   obtained  using  our  sample  of
  calibration lamps  (black dots).  The two CCDs  are shown separately
  (CCD1,  top; CCD2, bottom).   The blue  stars represent  the average
  value of the centers, which we take as reference for all lamps.  The
  red triangles show  the nominal center positions.  The  limits of the
  plots correspond to the parameter  space covered by the trial values
  used to determine the position  of the centers.  Positions are given
  in  binned  pixels on  the  chips.  Abscissas  larger than  1049  or
  negative imply centers outside the CCDs.}
\label{fig:center}
\end{figure}

Figure  \ref{fig:center}  shows the  best  centers  obtained for  each
calibration lamp image, and the mean value over our whole sample. They
turn out to be

\begin{displaymath}
(x_{\rm c1},y_{\rm c1})=(1053.6\pm 2.2, 979.5\pm1.8),
\end{displaymath}
%
for CCD1, and
%
\begin{equation}
(x_{\rm c2},y_{\rm c2})=(-9.8\pm2.7, 975.8\pm1.1),
\label{eq:center}
\end{equation}
%
for CCD2, where each center refers  to its own CCD.  The error bars in
Eq.~(\ref{eq:center})  represent  the  standard  deviation  among  the
values  obtained from  the  different rings,  therefore, assuming  the
errors of the different measurements to be independent, the error bars
of the average  values should be scaled down by a  factor of the order
of  the square  root of  the number  of rings  ($\sim$4,  e.g., Martin
1971).

In  the following  we use  the mean  values as  the centers  for every
calibration lamp.  The nominal values of the centers are also included
in Fig.~\ref{fig:center} as triangles.  The nominal center of CCD1 may
be marginally in agreement with our empirical determination.  However,
the center of CCD2 is clearly inconsistent with its nominal value.

%

\subsection{Calibration constants}
\label{sec:cc}

\begin{figure*}[!ht]
\centering
\includegraphics[width=0.8\textwidth]{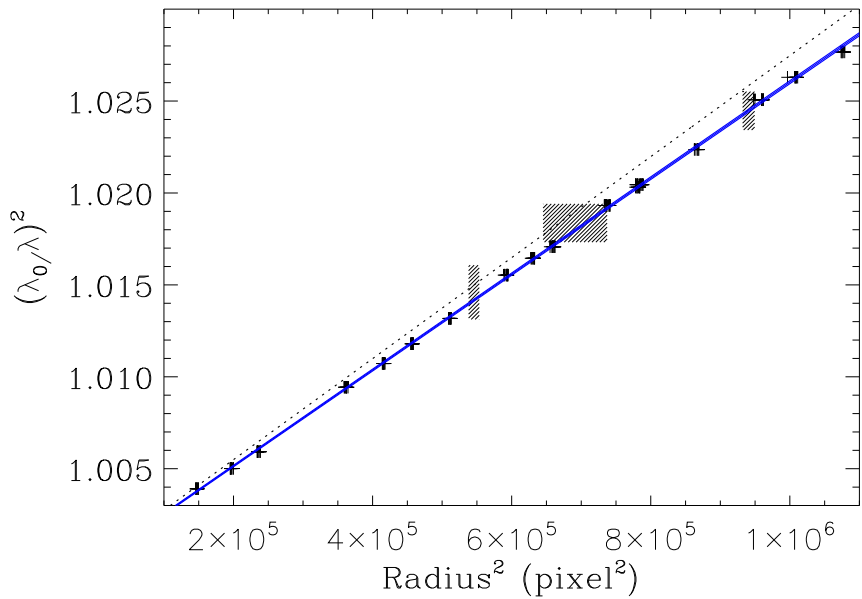}
\caption{Fit used to  derive empirically the value of  the camera lens
  effective focal length ($f_2$).   Black crosses represent the values
  measured using the  calibration lamps, one for each  region of every
  ring  image.   The  blue  line   represents  the  best  fit  to  our
  measurements. Hatched  areas indicate  the range of  possible values
  for the three sky rings.  The  black dotted line shows the result to
  be expected using the nominal value  for $f_2$, which is way off the
  observed values.  }
\label{fig:cal}
\end{figure*}

A trivial manipulation of Eq.~(\ref{eq:caleq1}) renders

\begin{equation}
\Big[{{\lambda(0)}\over{\lambda(R)}}\Big]^2=1+f_2^{-2}\,R^2,
\label{eq:cal3}
\end{equation}
%
thus,  a straightforward  linear regression  allows us  to  derive the
value of the  effective focal length of the  camera lens ($f_2$) given
the  positions  on  the  image  ($R$) of  lines  of  known  wavelength
($\lambda(R)=\lambda_{\rm line}$). Figure~\ref{fig:cal} shows the data
and  fit leading  to the  empirical determination  of $f_2$.   We have
measured  the distance  from the  center  of the  system (obtained  in
\S~\ref{sec:tfcenter}) to  the positions  of the rings  produced by
the    17     different    monochromatic    images     described    in
\S~\ref{sec:data}.  As  it was done  in the previous  section, this
distance was measured 8 times corresponding to the 8 different sectors
in   which   each   ring   was   divided.    They   are   plotted   in
Fig.  \ref{fig:cal},  showing  a  very  small  scatter  that  provides
confidence on the procedure.  The  best value for $f_2$ was derived by
fitting a  straight line to  the points in Fig.~\ref{fig:cal}  using a
standard linear least squares minimization routine.  Once the slope is
properly   transformed    assuming   1~binned   pixel    =   30~$\mu$m
(Table~\ref{tab:summary}), the best fit gives

\begin{equation}
f_2  = 185.70  \pm 0.17~{\rm mm}. 
\label{eq:focal}
\end{equation}
%
The fit also provides the $y-$intercept. It turns out to be consistent
with    one    ($0.99995\pm0.00006$)    supporting    the    use    of
Eq.~(\ref{eq:cal3}) to represent the wavelength variation along OSIRIS
FOV.

Note  how  the empirical  $f_2$  differs  from  the nominal  value  of
$181$~mm (Table~\ref{tab:summary}).  This  difference in the effective
focal length  of the camera  lens was not completely  unexpected. From
astrometry and laboratory tests we know that the actual plate scale of
OSIRIS  is  $\sim$0\farcs127~pixel$^{-1}$   rather  than  the  nominal
0\farcs125~pixel$^{-1}$ (Table~\ref{tab:summary}). This difference can
only be  accounted for  if the nominal  focal lengths differ  from the
actual values (see Eq.~(\ref{eq:scale})).

The  hatched areas in  Fig.~\ref{fig:cal} show  the position  of three
rings    from    sky    lines    also   measured    in    this    work
(\S~\ref{sec:data}).   The  measurements  are  not  represented  as
points  but as regions,  since they  are not  single lines  but series
covering  a fairly  broad spectral  range  (Table~\ref{tab:data}).  We
have  considered the  wavelengths of  the  sky lines  covered by  each
filter as  an error in  the ordinate axis.  In the abscissa  axis, the
errors correspond  to the position of  the 8 different  sectors in the
image. It is worth pointing out  that even if the chosen sky lines are
not suitable to be used  for wavelength calibration, their position on
OSIRIS  focal plane  are  fully consistent  with  the new  calibration
carried out in this work (Fig.~\ref{fig:cal}).

Once $f_2$ has been obtained, and in order to compare with the nominal
values, we have derived the  calibration constants $C^*$ and $C$ using
Eqs.~(\ref{eq:approx2}) and (\ref{eq:linking}),

\begin{displaymath}
C^*=14.499\pm0.027{\rm ~m}^{-2}= (3.2624\pm0.0057)\times 10^{-9}\, {\rm pixel}^{-2},
\end{displaymath}
\begin{equation}
C=(7.517\pm 0.014)\times 10^{-4}\,{\rm arcmin}^{-2}\,
[S/0\farcs125 {\rm\,pixel}^{-1}]^{-2}.
\label{eq:calconst}
\end{equation}
%
$C$  uses $S=  0\farcs125$\,~pixel$^{-1}$ for  the sake  of comparison
with the nominal values, although the wavelengths calibrated using $C$
in  Eq.~(\ref{eq:caleq2}) are  independent  of the  plate scale.   The
errors in $C$ and $C^*$  have been calculated by propagating the error
of $f_2$  (Eq.~(\ref{eq:focal})).  In the same way,  the ratio between
the  GTC  and  OSIRIS   collimator  effective  focal  lengths  can  be
determined  yielding $f_1/f_0  =  (7.5024\pm0.0070)\times10^{-3}$ with
$S= 0\farcs125$\,~pixel$^{-1}$.

\begin{figure}[!ht]
\centering
\includegraphics[width=0.7\textwidth]{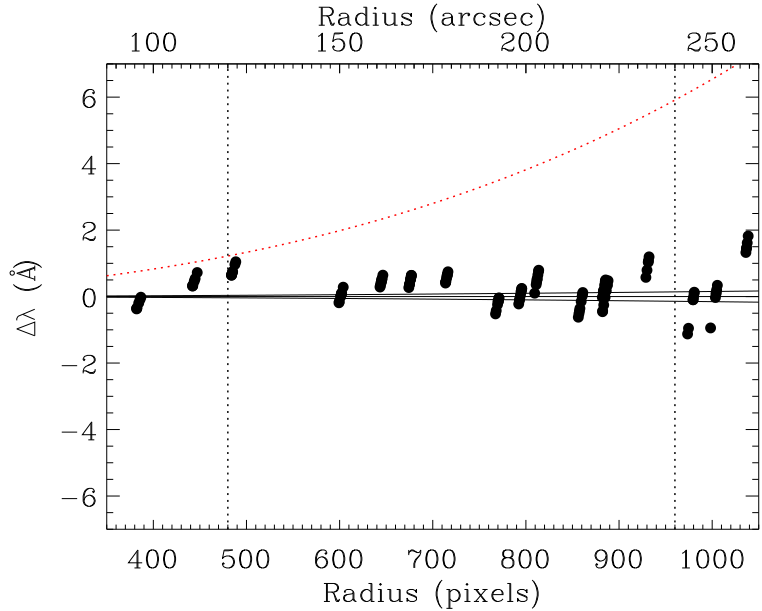}
\caption{Wavelength difference  in \AA\ between  the actual wavelength
  of the calibration lines and  that predicted by the new calibration.
  Black dots represent  the measurement for each ring  sector of every
  calibration lamp.  The black solid lines show the uncertainty caused
  by the error in $f_2$.   The red dotted line indicates the deviation
  from the actual wavelength of the calibration lines given by nominal
  calibration.  The scale on top has been calculated using the nominal
  plate  scale.   Vertical dotted  lines  show  the  radii of  the  MF
  ($\sim$2\arcmin) and the non-vignetted FOV ($\sim$4\arcmin) }
\label{fig:cal_OK}
\end{figure}

Figure~\ref{fig:cal_OK}  represents the  difference  between the  true
wavelengths of the  lines and that inferred using  our calibration. It
is equivalent to Fig.~\ref{fig:badcal}  but using the new calibration.
The reason  for including this  figure is threefold.  First,  it shows
how  the   systematic  trend  present   in  Fig.~\ref{fig:badcal}  has
disappeared,  meaning that  the new  calibration holds  throughout the
whole  OSIRIS FOV.   Second, the  observed scatter  provides  an upper
limit to the error in tuning the central wavelength of OSIRIS.  As one
can see  from Table~\ref{tab:data}, all  but one of  the monochromatic
images  were taken with  a different  $\lambda(0)$.  Any  random error
existing   in   setting   $\lambda(0)$   would   be   transferred   to
$\Delta\lambda$ as  a random fluctuation of amplitude  similar to that
of   $\lambda(0)$  (see   Eq.~(\ref{eq:needfor})).   The   scatter  in
Fig.~\ref{fig:cal_OK} does not exceed 1\AA, and so does the setting
of  $\lambda(0)$.    Finally,  Fig.   6   also  shows  $\Delta\lambda$
considering the original calibration (the red dotted line).  The error
remains within  the acceptable limit of $\Delta\lambda  < 1~$\AA\ when
$R < 2$\arcmin , which approximately coincides with OSIRIS MF.

\subsection{Error budget}
\label{sec:budget}

Three main  sources of uncertainty burden  the wavelength calibration,
namely, errors in the central wavelength $\delta\lambda(0)$, errors in
the  calibration  constant $\delta  C^*$,  and  errors  in the  center
position $\delta x_{\rm c}$, and  $\delta y_{\rm c}$. The error in the
central  wavelength  has  not  been  explicitly  treated  before  but,
according to the arguments given in the previous section, it is within
the nominal 1\,\AA\,  set by the procedure to  tune OSIRIS (see OSIRIS
manual).

Using Eq.~(\ref{eq:caleq1}), it is  possible to propagate these errors
onto the wavelength calibration $\delta\lambda(R)$. The result of this
exercise turns out to be
%
\begin{equation}
\frac{\delta\lambda(R)}{\lambda(R)}=\frac{\delta\lambda(0)}{\lambda(0)},
\end{equation}
%
for the central wavelength error $\delta\lambda(0)$,
%
\begin{equation}
\frac{\delta\lambda(R)}{\lambda(R)}=\frac{-\delta C^*}{2\,C^*}
\,\left[1-\frac{\lambda^2(R)}{\lambda^2(0)}\right],
\end{equation}
%
for the calibration constant error $\delta C^*$, and finally,
%
\begin{equation}
\frac{\delta\lambda(R)}{\lambda(R)}=\frac{\delta x_{\rm c}\cos\phi}{R}
\,\left[1-\frac{\lambda^2(R)}{\lambda^2(0)}\right],
\label{eq:err3}
\end{equation}
%
for the error of the center in the $x$-axis ($\delta x_{\rm c}$).  The
variable $\phi$  stands for the azimuth  of the point  relative to the
center,  and  it  has  been  introduced for  convenience  since  using
$|\cos\phi|=1$  in  Eq.~(\ref{eq:err3})   grants  an  upper  limit  to
$\delta\lambda$.   The  error due  to  uncertainties  in the  vertical
direction is  formally identical to that for  the horizontal direction
replacing $\delta x_{\rm c}\cos\phi$ with $\delta y_{\rm c}\sin\phi$.

\begin{figure}[!ht]
\centering
\includegraphics[scale=0.7]{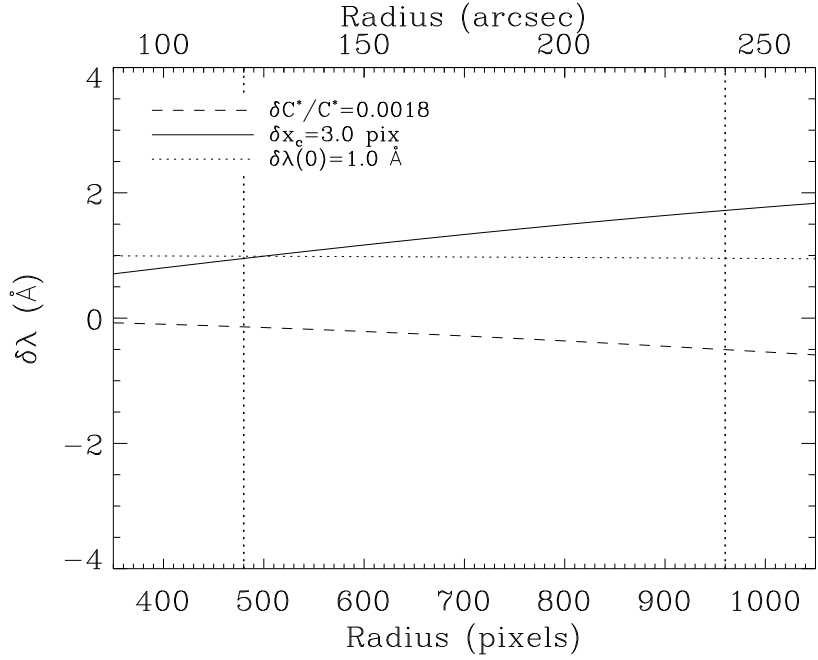}
\caption{Wavelength errors  in the calibration  ($\delta\lambda$) as a
  function of the  distance to the center of the  FOV. As indicated in
  the  inset,  the  curves  represent  a  mis-tuning  of  the  central
  wavelength  ($\delta\lambda(0)\not=  0$,  dotted  line),  errors  in
  calibration   constant  ($\delta   C^*\not=0$,  dashed   line),  and
  uncertainties locating  the center of the FOV  ($\delta x_c\not= 0$,
  solid line).   The errors we consider are  realistic, except perhaps
  for an  overestimate of $\delta  x_c$ (see main text).   The central
  wavelength  was assumed to  be tuned  to H$\alpha$.  Vertical dotted
  lines  show   the  radius  of   the  MF  ($\sim$2\arcmin)   and  the
  non-vignetted FOV ($\sim$4\arcmin). }
\label{fig:error_budget}
\end{figure}

In order to illustrate the  relative importance of the various sources
of    error,    Fig.~\ref{fig:error_budget}    shows    the    typical
$\delta\lambda(R)$ to  be expected with our  calibration, assuming the
central wavelength  to be tuned  to H$\alpha$ (6563~\AA).   The dotted
line  corresponds   to  the  effect   of  1\AA\  uncertainty   in  the
determination of the central  wavelength.  It sets a minimum threshold
to  the calibration error  in the  full OSIRIS  FOV.  The  dashed line
considers $\delta C^*=0.027$~m$^{-2}$,  which corresponds to the error
in  $f_2$  derived  in   our  calibration  (see  \S~\ref{sec:cal}  and
Table~\ref{tab:summary}).   The  calibration   constant  seems  to  be
precise enough and does not limit the wavelength calibration. Finally,
the solid  line corresponds to  a center uncertainty given  by $\delta
x_{\rm c}\,\cos\phi=3$~pixels.  It produces an error in the wavelength
scale that is almost 2\,\AA\ in  the outer parts of the FOV, exceeding
the threshold set by the central wavelength.  Three pixels represent a
significant  overestimate  of  the  uncertainties we  expect  for  the
centers;  if we  consider a  more realistic  $\delta x_c\,\cos\phi\leq
1.5$~pixels, then  $\delta\lambda(R)\leq 1$\AA,  and the error  in the
center  position  does  not   limit  the  calibration.   We  use  this
overestimated  value to  illustrate  that {\em  the  precision in  the
  center  turns  out  to  be  the critical  point  of  the  wavelength
  calibration.}

The  question  arises   as  to   whether  $\delta\lambda(R)\leq
  1\,$\AA\ suffices to compute the line fluxes that triggered our work
  (\S~\ref{sec:introduction}).  We address the problem assuming the TF
  bandpass           to          behave           as          expected
  \citep[e.g.,][]{bornwolf80,beckers98},
%
\begin{equation}
\tau=\left[1+\left[\frac{2\,\left(\lambda_s-\lambda(R)\right)}{\sigma_\lambda}\right]^2\right]^{-1},
\label{eq:trans_tau}
\end{equation}
%
with $\sigma_\lambda$  the FWHM of  the TF transmission  bandpass, and
$\lambda(R)$  the wavelength  calibrated above.   Consider  a spectral
line of wavelength $\lambda_s$ and integrated flux $F$. Then, the flux
measured in a given point of  the focal plane $R$, $F_m(R)$, turns out
to be
%
\begin{equation}
F_m(R)=\tau\,F,
\label{eq:flux_law}
\end{equation}
%
where  we have  assumed  the line  to  be much  narrower  than the  TF
transmission  (as it  is indeed  the case  for the  emission  lines in
typical  \HII\, regions).   Equation~(\ref{eq:flux_law})  provides the
recipe to  retrieve the  true flux from  the observed one  knowing the
transmission.  If  $\lambda(R)$ is uncertain then  $\tau$ is uncertain
as well,  producing the  error on  $F$ that we  try to  evaluate.  The
error in the retrieved flux $\delta F$ due to the error in the central
wavelength  $\delta\lambda(R)$  was  derived   by  applying  the  law  of
propagation      of      errors      \citep[e.g.,][]{martin71}      to
Eq.~(\ref{eq:flux_law}), and it is given by
%
\begin{equation}
\frac{\delta F}{F}\simeq 
-\frac{d\ln\tau}{d\lambda(R)}\delta\lambda(R).
\label{eq:err_flux_law}
\end{equation} 
%
After some trivial manipulations, the previous equation becomes
%
\begin{equation}
\frac{\delta F}{F}\simeq 
-4\,\big[(1-\tau)\,\tau\big]^{1/2}\,
\frac{\delta\lambda(R)}{\sigma_\lambda}.
\label{eq:my_this_eq}
\end{equation} 
%
Since $0\le \tau\le 1$, then $\big[(1-\tau)\,\tau\big]^{1/2} \le 0.5$,
and so
%
\begin{equation}
\Big|\frac{\delta F}{F}\Big|\leq
2\Big|\frac{\delta\lambda(R)}{\sigma_\lambda}\Big|.
\label{eq:upper_limit}
\end{equation} 
%
For    $\delta\lambda(R)\leq    1\,$\AA\   and    a    TF   FWHM    of
$\sigma_\lambda\simeq    20$\,\AA,   Eq.~(\ref{eq:upper_limit})   yields
$|\delta  F/F|\leq  0.1$.   This   error  is  small  enough  for  most
applications,  including those mentioned  in \S~\ref{sec:introduction}
-- e.g., it  provides SFRs within  10\,\% using the  customary recipes
\citep[e.g.,][]{kennicutt98},  and   it  is  much   smaller  than  the
intrinsic scatter  of the sulphur  doublet ([SII]$\lambda\lambda$6717,
6731)       used      to       determine       electron      densities
\citep[e.g.,][]{bresolinkennicutt02}.   Three  final  comments are  in
order.   First, a  scaling factor $\tau$  does not  affect our  error
estimate,  which   is  the  reason   why  the  peak   transmission  in
Eq.~(\ref{eq:trans_tau})     was      set     to     one.      Second,
Eq.~(\ref{eq:flux_law})  implicitly assumes  the observed  spectrum to
have no  continuum. If there  is continuum, then the  arguments remain
valid  replacing  $F_m(R)$  with  the difference  between  the  fluxes
measured  in line  and  in  continuum.  Finally,  the  upper limit  in
Eq.~(\ref{eq:upper_limit}) significantly  overestimates the true error
when the  wavelength setting  is close to  the peak  transmission (see
Eq.~(\ref{eq:my_this_eq}) with $\tau\simeq 1$).

\section{Wavelength calibration protocol}
\label{sec:protocol}

Based in our  previous experience, this section puts  forward a set of
guidelines  to carry  out an  accurate wavelength  calibration  of the
upcoming  scientific  data.    It  involves  routine  observations  of
calibration  lamps together  with  the astronomical  images. Since  no
extensive  tests  on  the  stability  of  the  OSIRIS  TF  center  and
calibration  constant have  been  performed yet,  they  are needed  to
ensure  the accuracy  presented in  this  work and  to identify  other
possible errors.  The  protocol will also allow to  study and quantify
the wavelength dependence of the calibration parameters.

In general,  the calibration process  should consist of  the following
four steps:

\begin{enumerate}
\item  Obtain   a  number  of  calibration  lamp   images  during  the
  observation  run  covering  the   full  FOV.  Preferably,  select  a
  monochromatic emission  line from a  lamp with a wavelength  near to
  the  scientific observation.   Then, map  its positions  along the
  OSIRIS FOV by changing the central wavelength of the TF.
\item Calculate the center of the system using the rings corresponding
  to the monochromatic lines. Check that it is the same for all rings
  within a reasonable error (a few pixels; see Fig. \ref{fig:center}).
\item Calculate the value of  the effective focal length of the camera
  lens ($f_2$) by fitting linearly Eq.~(\ref{eq:cal3}).  This equation
  should be  fitted in pixels  or in a  physical length (mm)  to avoid
  crosstalk with errors in  the plate scale.
\item  Calibrate the  scientific  images by  assigning the  wavelength
  given by Eq.~(\ref{eq:caleq1}) to  every pixel.  Use the empirically
  derived $f_2$ and optical center.
\end{enumerate}

We  recommend to  apply the  calibration  described here  to each  CCD
independently,  and before applying  astrometric corrections  to avoid
other error  sources. Eventually,  if the instrument  turns out  to be
stable, the calibration will consist only of step 4 with $f_2$ and the
center derived in this paper.

For observations already in hand using the full OSIRIS FOV, we suggest
to recalibrate the images  using the prescriptions and constants given
in   this   paper   (Eqs.~(\ref{eq:caleq1}),  (\ref{eq:center}),   and
(\ref{eq:focal})).   Whenever  a  calibration  ring  is  available  as
ancillary  data, we  recommend computing  the center  and  compare its
value  with  our  estimate  (Eq.~(\ref{eq:center})).  A  remainder  is
important: as we  explained in \S  5.1, the  trials for determining
the best center have been carried out  on the raw data. If the data to
be re-calibrated have already been processed (e.g., overscan removed),
then the values of the best center have to be shifted accordingly.

{\em  Existing observations  using only}  OSIRIS MF  {\em do  not need
  re-calibration.}  The  approximate original prescriptions  in OSIRIS
manual suffice.

\section{Conclusions}
\label{sec:conclusions}

OSIRIS TF and GTC represent a unique combination to study the physical
properties  of extended  astronomical  sources.  They  are capable  of
providing high S/N, narrow band  images, over a relatively large field
with  an excellent  spatial sampling.   The OSIRIS  TF is  based  on a
conventional  FPI mounted  in collimated  beam. FPI  based instruments
provide  an area  of the  FOV,  the MF,  where the  wavelength can  be
considered constant within the  band-pass of the TF.  However, outside
the MF there is a significant shift in wavelength which depends on the
distance to  the optical  axis.  Therefore, in  order to  benefit from
OSIRIS large FOV, an accurate wavelength calibration is needed.

Motivated by inconsistencies in the measurements of line ratios during
an observational campaign, we  decided to re-calibrate empirically the
equation which describes the shift  in wavelength with position on the
FOV.   Using spectral calibration  lamps, we  found that  the proposed
nominal calibration suffices only in  the central 2\arcmin of the FOV,
but  fails elsewhere.  The  approximate equation  originally given  to
describe the  wavelength shift  with radius is  not correct  for large
radii,    and   the    exact   equation    must   be    used   instead
(Eq.~(\ref{eq:caleq1})).

We  identify  three  different  sources  of error  in  the  wavelength
calibration,    namely,    errors    in   the    central    wavelength
$\delta\lambda(0)$, errors  in the calibration  constant $\delta C^*$,
and  errors in  the center  position  $\delta x_{\rm  c}$ and  $\delta
y_{\rm c}$.  The  error in the central wavelength  is intrinsic to the
procedure used  to tune the  instrument at the telescope.   We confirm
that the  errors introduced  in this process  are $\lesssim  1$\AA, as
stated  in the  instrument manual.   The calibration  constant  in the
procedure proposed here only depends  on the effective focal length of
the  camera lens.   It has  been empirically  determined with  a small
error (Eq.~(\ref{eq:focal})),  good enough to  discard the calibration
constant as a source of uncertainty.  We have also found a new optical
center    defined   independently    for   the    two    OSIRIS   CCDs
(Eq.~(\ref{eq:center})) which  improves significantly the calibration.
Actually, the position  of the center seems to  be the factor limiting
the  final  accuracy.   Considering  all  uncertainties,  the  updated
calibration provides an accuracy of $\Delta\lambda \lesssim 1$\AA\, in
the entire OSIRIS FOV.

The  use  of  the  new  calibration makes  OSIRIS  ideal  for  studies
involving  spectrophotometric measurement  of large  galaxies covering
the  full FOV,  such as  those that  triggered the  present  work (see
\S~\ref{sec:introduction}).   However,  it  is  of  broader  interest.
Whatever  study  using  the  entire  FOV will  benefit  from  the  new
calibration (e.g.,  extended planetary  nebulae, clusters of  stars or
galaxies distributed in a large area, etc).

Based in our previous experience, a wavelength calibration protocol is
proposed  in  \S~\ref{sec:protocol}.   It  will  allow  to  check  the
stability of both OSIRIS center and the calibration constant, i.e., to
see  whether  they vary  over  long  periods,  when OSIRIS  optics  is
re-aligned,  or  if different  wavelength  ranges  are  used.  In
  particular, when the blue etalon will be at work, a similar study on
  the wavelength calibration must be done.

\acknowledgments 
Thanks are due  to the GTC staff, in  particular to A. Cabrera-Lavers,
for their support and for providing the data used in this study.
Discussions   with  J.~L.~Rasilla,   R.   L\'opez,   and  J.~S\'anchez
Capuchino, opticians involved in  the design and integration of OSIRIS
and  ICM/GTC, were  extremely  clarifying.  We  are  also indebted  to
A. Manescau and J.~ A.~ Bonet for support and advice.
This  work has  been partially  funded  by the  Spanish MICINN  (grants
AYA2007-67965-C03-01     and     AYA2010-21887-C04-04)     and     the
Consolider-Ingenio Program CSD2006-00070.


\end{document}